\documentclass[%
 aip,
 amsmath,amssymb,
 reprint,%
]{revtex4-1}

\usepackage{graphicx}
\usepackage{dcolumn}
\usepackage{bm}
\usepackage[utf8]{inputenc}
\usepackage[T1]{fontenc}
\usepackage{mathptmx}
\usepackage{etoolbox}
\usepackage[version=3]{mhchem} 
\usepackage{tabularx}
\usepackage{booktabs}
\usepackage{braket}
\usepackage{multirow}
\usepackage{multirow}
\usepackage{stfloats}
\usepackage{tikz}
\usepackage{amsmath}
\usepackage{mciteplus}
\usepackage{subcaption}
\usepackage{float}

\makeatletter
\def\@email#1#2{%
 \endgroup
 \patchcmd{\titleblock@produce}
  {\frontmatter@RRAPformat}
  {\frontmatter@RRAPformat{\produce@RRAP{*#1\href{mailto:#2}{#2}}}\frontmatter@RRAPformat}
  {}{}
}%
\makeatother
\begin{document}

\preprint{AIP/123-QED}
\title[\textcolor{black}{Universal Approach for Determining Multi-Dimensional Anharmonic Vibrations from Electronic Quantum Methods}]{\textcolor{black}{Universal Approach for Determining Multi-Dimensional Anharmonic Vibrations from Electronic Quantum Methods}}
\author{Kushantha P. K. Withanage}
\email{kpwithanage@utep.edu}
\affiliation{Department of Physics, the University of Texas at El Paso, El Paso, TX-79968, United States.}
\author{Jesús Pedroza-Montero}
\affiliation{Department of Physics, the University of Texas at El Paso, El Paso, TX-79968, United States.}
\author{ M.F. Islam}
\affiliation{Department of Physics, the University of Texas at El Paso, El Paso, TX-79968, United States.}
\author{ Koblar A. Jackson}
\affiliation{Department of Physics, Central Michigan University, MI-48859, United States}
\author{Mark R. Pederson}
\affiliation{Department of Physics, the University of Texas at El Paso, El Paso, TX-79968, United States.}
\date{\today}

\begin{abstract}
  We present a simple and efficient method to incorporate anharmonic effects in the vibrational \textcolor{black}{analyses} of molecules within density functional theory (DFT) calculations. This approach is closely related to the traditional vibrational \textcolor{black}{configuration} interaction (VCI) technique, which uses the harmonic oscillator wavefunctions as the basis. In our implementation, we employ Gaussian-type orbitals (GTOs), with polynomial prefactors, as the basis set to evaluate the anharmonic Hamiltonian. Although these basis functions are non-orthogonal, the matrix elements such as overlap, kinetic energy terms, and position moments can be evaluated analytically. The terms in the Hamiltonian due to the anharmonic potentials are numerically calculated on a Hermite-Quadrature grid. The potentials can be evaluated using any electronic structure method. This framework enables us to accurately calculate the anharmonicity-corrected vibrational frequencies, the fundamental frequencies, and the corrections to bond lengths in diatomic molecules. This method is also generalized to handle coupled anharmonic oscillators, which is essential to model more complex phenomena such as nitrogen tunneling in the umbrella mode of ammonia (NH$_3$) and Fermi resonances in carbon dioxide (CO$_2$).   
\end{abstract}

\maketitle

\section{Introduction}  
Molecular vibration is central to determining how molecules interact with electromagnetic radiation, particularly in the infrared (IR) region. These vibrations are quantized. They yield characteristic energy levels that serve as a unique identifier for each molecule. This sets the basis for infrared and Raman spectroscopy in chemistry and materials science, where characteristic vibrational frequencies allow us to determine the types of atoms present and the molecular structure\cite{Atkins2010,Herzber46}. In addition to spectroscopy, molecular vibrations are important for understanding other important phenomena. For example, molecules such as carbon dioxide (CO$_2$), methane (CH$_4$), and water vapor (H$_2$O) possess IR-active vibrational modes that absorb outgoing thermal radiation from Earth’s surface. The absorbed energy is then re-emitted, trapping heat within the atmosphere. A particularly well-known example is the Fermi resonance in CO$_2$, which enhances IR absorption around 15 $\mu$m, a critical region for Earth’s heat emission. This behavior is a key factor in the contribution of greenhouse gases to global warming\cite{wordsworth2024fermi,change2007climate}. Vibrational excitation also influences molecular reaction dynamics and the energy transfer between atoms/molecules affecting  reaction rates and reaction pathways\cite{crim2008chemical}. For these reasons, the ability to simulate and accurately predict molecular vibrations and their effects is of great interest in both fundamental and applied sciences. 
In this paper, we present a method that can be used to accurately evaluate the vibrational frequencies of molecules beyond the harmonic approximation. \textcolor{black}{This method is based on a single-center Gaussian basis set. The use of single-center localized Gaussian basis sets to calculate harmonic and anharmonic vibrational frequencies is not new for either one- or N-dimensional vibronic wavefunctions. Perhaps the nuclear-electronic orbital (NEO) approach\cite{NEOSHS,SHSXLi}, for calculating vibrational frequencies of molecules, is the most established method that simultaneously determines both vibrational and electronic quantum effects. This paper offers a means for systematic identification of the physically interesting coupled vibrational modes in a multi-atom system and is viewed as a precursor to further departures from the use of single-center gaussian basis sets as the "go-to" basis for understanding the role of vibration in molecular systems. We acknowledge that this may be an ambitious task since the use of such functions for electronic structure methods have survived challenges from many suggestions for next-generation basis functions.}

In the harmonic approximation, the potential energy surface (PES) along each normal mode is assumed to be purely quadratic. However, real molecular vibrations deviate from this idealized behavior, resulting in anharmonicity. Accounting for anharmonicity not only shifts the computed vibrational frequencies closer to experimentally observed values, but also provides a more physically accurate description of molecular behavior, including energy level spacing, mode coupling, and vibrational intensity patterns\cite{Tomica_anh,barone2014fully}. When incorporating anharmonic effects in electronic structure calculations, the most widely used methods include vibrational perturbation theory (VPT2) \cite{barone2005anharmonic,10.1063/1.3695210}, vibrational self-consistent field (VSCF)\cite{bowman1986self,gerber1988self}, and vibrational configuration interaction (VCI) approaches \cite{christiansen2007vibrational,rauhut2007configuration,doi:10.1021/acs.jctc.9b00294}. The method presented here is equivalent to the VCI method. We will discuss this further in the theory and implementation section. 

In the following section, we introduce our framework and discuss its implementation. Then we apply the method to evaluate the anharmonic frequencies of diatomic molecules and corrections to the bond lengths due to the anharmonicity. To demonstrate the versatility of the method for coupled-anharmonic oscillators, we 
model nitrogen tunneling in the umbrella inversion mode in NH$_3$ and the Fermi-resonance phenomena in the CO$_2$ molecule. This is followed by a summary.

\section{Theory and Implementation}

First, we introduce how we incorporate anharmonicity for diatomic molecules. The PES along a normal mode can be expanded as a polynomial of the displacement (x) about the equilibrium, as in Eq.~\ref{PES}.  When $m>2$, the PES is anharmonic.   
\begin{equation}
    V(x)= \sum_{i=0}^m a_ix^i
    \label{PES}
\end{equation}
The Hamiltonian of the anharmonic oscillator is
\begin{equation}
   \hat H=-\frac{1}{2M'}\frac{d^2}{dx^2}+V(x)
\end{equation}
where $M'$ is the effective mass of the mode. In the VCI method, the quantum harmonic oscillator wavefunctions including higher excited states are used as basis functions for this Hamiltonian. However, we choose to use the simpler product functions shown in equation \textcolor{black}{\ref{basis}} as our basis functions. These span the same space as the wave functions of the quantum harmonic oscillator.   

\begin{equation}
\psi_n(x)= x^ne^{-A^2x^2/2}
\label{basis}
\end{equation}
In this expression $A^2=\omega/M'$ and $\omega$ is the frequency associated with the harmonic approximation or it can alternatively be used as a variational parameter for cases that strongly deviate from the harmonic idealization. 
As a result, we 
work with a non-orthogonal basis set. This results in non-zero off-diagonal matrix elements in the overlap matrix, $s_{n,m}$. The integrals to be evaluated have the form:

\begin{equation}
    s_{n,m}=\braket{\psi_n|\psi_m}=\int_{-\infty}^{+\infty}x^{n+m}e^{-A^2x^2}dx
    \label{over}
\end{equation}
The overlap matrix elements can be calculated analytically using recursion relations. 

Similarly, the contribution from the kinetic energy operator to the Hamiltonian matrix elements, $t_{n,m}$ can also be evaluated analytically using recursion relations. The required integrals are:

\begin{align}
    t_{n,m}&=\frac{-1}{2M'}\braket{\psi_n|\frac{d^2}{dx^2}|\psi_m} \nonumber \\ 
    &=\frac{-1}{2M'}\int_{-\infty}^{+\infty}x^{n}e^{-A^2x^2/2}\frac{d^2[x^{m}e^{-A^2x^2/2}]}{dx^2}dx   \nonumber \\
    &=\textcolor{black}{\frac{-1}{2M'}\Big[m(m-1)s_{n,m-2}-A(2m+1)s_{n,m}+A^2s_{n,m+2}\Big]}
    \label{kin}
\end{align}

The only term 
in the anharmonic Hamiltonian 
that must be evaluated numerically is the contribution from the anharmonic potential: 
\begin{equation}
   <\psi_n|V(x)|\psi_m >= \int_{- \infty}^{+\infty} x^{n+m}e^{-A^2x^2}V(x)dx
   \label{4}
\end{equation}

For the integration in equation \ref{4},  we utilize the Gauss–Hermite quadrature method. The number of grid points ($N$) is chosen to accurately perform the matrix element in equation \ref{4} with highest order of $x$ in $f(x)=x^{n+m}V(x)$. Here it is emphasized that the quadrature approach removes the need to numerically extract the expansion coefficents $\{a_i\}$ in Eq.~\ref{PES}. Given the weights $w_i$ for the grid, the matrix elements are reduced to:

 \begin{equation}
      <\psi_n|V(x)|\psi_m >=\sum_i^Nw_ix_i^{n+m}V(x_i)
 \end{equation}
 However, Eq.~\ref{4}  can also be rewritten as the sum of two integrals by using an integration by parts. 
In each integral, the order of the integrand is 
reduced by 2. Due to this, a smaller number of Hermite-quadrature points is required for the numerical integrations compared to Eq.~\ref{4}. The NRLMOL DFT software package \cite{pederson1990variational,jackson1990accurate} used for this study has the ability to accurately and efficiently calculate the gradients of energy with respect to atomic positions. 

\begin{widetext}
\begin{equation}
   <\psi_n|V(x)|\psi_m >=  \frac{1}{2A^2}\Big[\int_{-\infty}^{+\infty}\frac{dV}{dx}x^{n+m-1}e^{-A^2x^2}dx + \int_{-\infty}^{+\infty}(n+m-1)V(x)x^{n+m-2}e^{-A^2x^2}dx\Big]
\end{equation}
\end{widetext}

In addition to computing vibrational energies, our method allows us to evaluate properties such as corrections to bond lengths 
analytically. 
The correction to the bond length for $k^{th}$ state is simply given by:    
\begin{align}  
\braket{x}_k &=\sum_{ij}C^k_nC^k_m\braket{\psi_n|x|\psi_m} \\ \nonumber
&=\sum_{nm}C^k_nC^k_m\int_{-\infty}^{+\infty} x^{n+m+1}e^{-A^2x^2}dx\\ \nonumber
&=\sum_{nm}C^k_nC^k_ms_{n,m+1} \\ \nonumber
 \mathrm{or} \notag \\  \nonumber
&=\sum_{nm}C^k_nC^k_ms_{n+1,m}
\end{align}
where $C^k_n$'s are the eigenvector coefficients of the solutions to the anharmonic Hamiltonian $\hat H$.
The higher order moments can also be evaluated similarly. 

So far we have described the use of our method for a 1-dimensional anharmonic oscillator, which is ideal for diatomic systems. In the next section, we discuss how our method can be extended to treat coupled-anharmonic vibrations in molecules with many atoms. 

\subsubsection{Coupled-anharmonic oscillators}
Let us assume that two vibrational modes of a molecule are coupled due to the anharmonicity in the PES. Then the PES is a function of two independent displacements ($V(x,y)$). In this case, a general basis function for the coupled, two-mode anharmonic oscillator Hamiltonian 
can be defined as the product of basis functions used for the single anharmonic oscillator problem discussed above. The general two-mode product basis function for coupled anharmonic oscillator is given by: 
\begin{equation}
\phi_n(x,y)=\psi^x_i(x)\psi^y_j(y)= x^ie^\frac{-A^2x^2}{2}y^je^\frac{-B^2y^2}{2}
\end{equation}
Therefore, the overlap between two coupled-basis functions can be evaluated as   
\begin{equation}
\braket{\phi_n|\phi_{n'}}=\braket{\psi^x_i|\psi^x_{i'}}\braket{\psi^y_j|\psi^y_{j'}}=S^x_{i,i'}S^y_{j,j'}
\label{c_over}
\end{equation}
The contribution to the Hamiltonian from the kinetic energy operator becomes,
\begin{equation}
    \braket{\phi_n|\frac{1}{2M_1}\frac{d^2}{dx^2}+\frac{1}{2M_2}\frac{d^2}{dy^2}|\phi_{n'}}=T^x_{i,i'}S^y_{j,j'}+S^x_{i,i'}T^y_{j,j'}
    \label{c_kin}
\end{equation}
Again, 
the expressions in equations \ref{c_over} and \ref{c_kin} can be evaluated analytically, as mentioned in the previous section. Similarly, the remaining term needed for construction of the Hamiltonian matrix elements is due to the potential (Eq. \ref{2D}).  This can be evaluated according to: 
\begin{align}
\label{2D}
\braket{\phi_n|V(x,y)|\phi_{n'}}&=\int_{-\infty}^{\infty}dx\int_{-\infty}^{\infty}dy V(x,y)x^{i+i'}e^{-A^2x^2}y^{j+j'}e^{-B^2y^2}.
\end{align}
To numerically 
evaluate the integrals in Eq.~\ref{2D}, we apply the Hermite-quadrature method. This is achieved by defining a 2-dimensional Hermite grid, which is formed from 1-dimensional Hermite grids established for the two uncoupled modes. In the case of coupling only two modes, the total number of basis functions to construct the Hamiltonian is $N^2$ where $N$ is the number of basis functions used per uncoupled mode. 
One could incorporate gradient information of $V(x,y)$ for the integral (Eq.~\ref{2D}). However, it turns out that, without at least second-order derivatives of the PES, the number of Hermite-quadrature points cannot be reduced in this two-dimensional numerical integration. Therefore, only the energies were used for this. 
This approach can be easily generalized to solve the Hamiltonian for m-coupled anharmonic oscillators, which would result in $N^m$ basis functions. 

\section{Results and discussion}
\subsection{Diatomic molecules}
\begin{figure*}[htb!]
    \centering
    \includegraphics[scale=0.5]{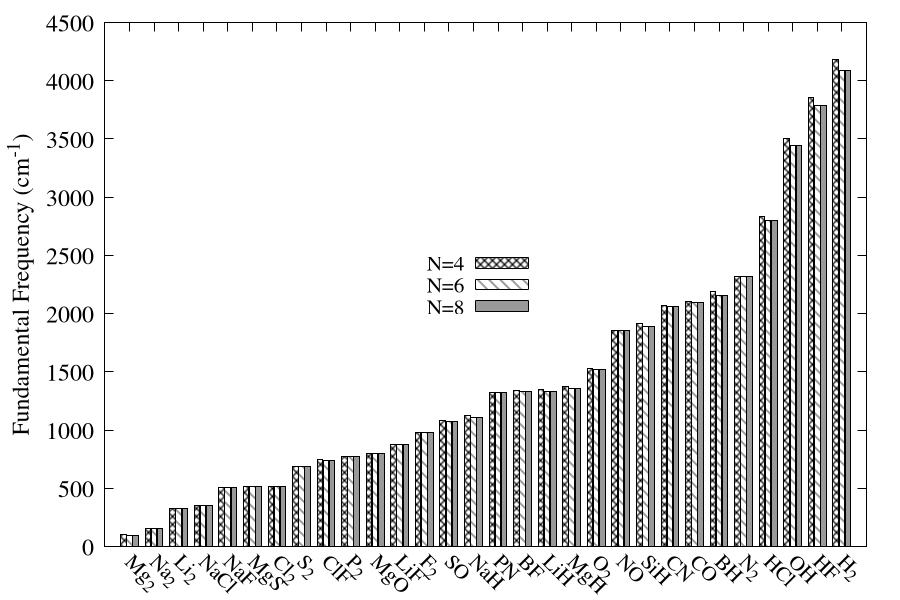}
    \caption{Fundamental frequencies of selected diatomic molecules calculated using different basis set sizes for the anharmonic Hamiltonian.}
    \label{anh_f_fund}
\end{figure*}

\begin{figure*}[htb!]
    \centering
    \includegraphics[scale=0.5]{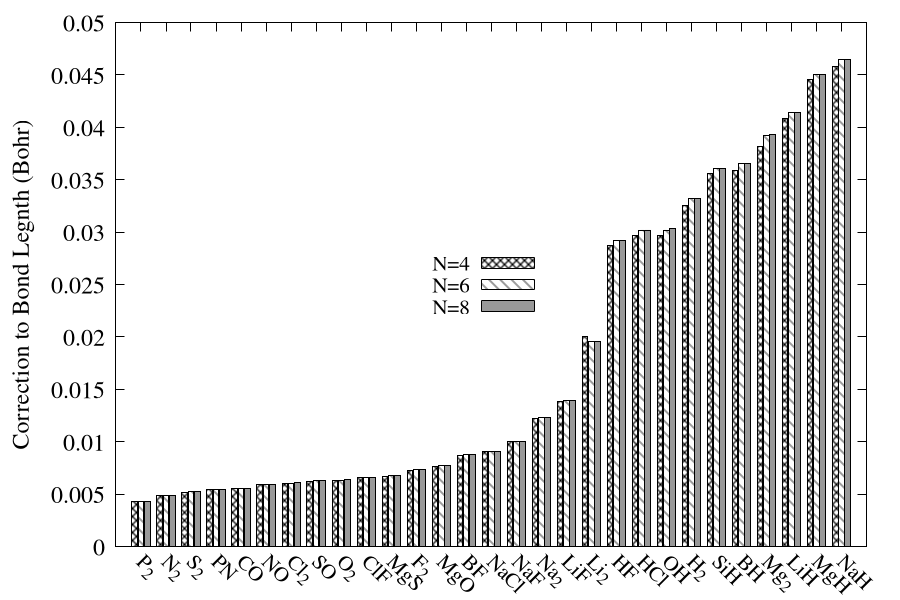}
    \caption{Anharmonic corrections to the bond lengths of selected diatomic molecules using different basis set sizes for the anharmonic Hamiltonian.}
    \label{bond_cor}
\end{figure*}

As our first application, we used our approach to evaluate the effect of anharmonicity in diatomic molecules. For this purpose, a set of diatomic molecules and radicals composed of atoms up to Cl was used. Here, we calculated the fundamental frequency, the anharmonicity-corrected frequency (i.e., the lowest energy vibrational transition of a molecule, typically the transition from the ground vibrational state ($\nu$=0) to the first excited vibrational state ($\nu$=1)) 
and the correction to bond length. We investigated the dependence of these properties on the size of basis set employed in constructing the anharmonic Hamiltonian. In Fig.~\ref{anh_f_fund}, we present the fundamental frequencies of the diatomic systems. Fig.~\ref{bond_cor} presents the correction to bond lengths due to the anharmonicity (refer to the supplementary materials for the complete dataset). We utilized 6, 8 and 10 Hermite grids points for N=4, N=6, and N=10  calculations, respectively. For the properties mentioned above, it can be seen that 6 basis functions (N=6) were sufficient to obtain converged results.\\ 

While the corrections to bond lengths are relatively small in diatomic molecules with strong covalent bonds, the largest corrections are observed in diatomics containing hydrogen and alkali or alkaline earth metal atoms(H-X). These H-X molecules typically have longer bond lengths, and it is well known that their potential energy surfaces (PES) show significant deviations from the harmonic approximation. However, when we repeated the calculations for NaH and HCl with deuterium and tritium, the corrections to the bond length are reduced as $\sim1/m^{1/2}$ where m is the mass of the hydrogen isotope used (see Table \ref{isotopes}). We also calculated the ratio between the lowest energy eigenvalues from the anharmonic and harmonic calculations($E_\mathrm{anh}$/$E_\mathrm{harm}$). The ratios are nearly same for hydrogen, deuterium, and tritium. This shows that the effect due to the anharmonicity on the energy (or frequency) remains the same (only a weak dependence on the hydrogen isotope mass).    
 
\begin{table*}[htb!]
    \centering
    \begin{tabular}{lccc}
    \toprule
          &   $^1$H & $^2$H & $^3$H  \\
         \midrule
        NaH  & 0.046 (0.977)&  0.032 (0.979)& 0.026 (0.979)\\
        HCl & 0.033 (0.988) & 0.021 (0.991)& 0.017 (0.992) \\
        \bottomrule
    \end{tabular}
    \caption{Correction to bond length (in Bohr) for different hydrogen isotopes. The corresponding $E_\mathrm{anh}$/$E_\mathrm{harm}$ ratios are given in parentheses.}
    \label{isotopes}
\end{table*}

\subsection{Coupled-anharmonic oscillator}

\subsubsection{Ammonia (NH$_3$): coupling between the umbrella mode and the symmetric stretch mode}
\begin{figure*}[htb!]
\centering
\begin{tikzpicture}
\node (D) at (4,-5) {\includegraphics[scale=0.6]{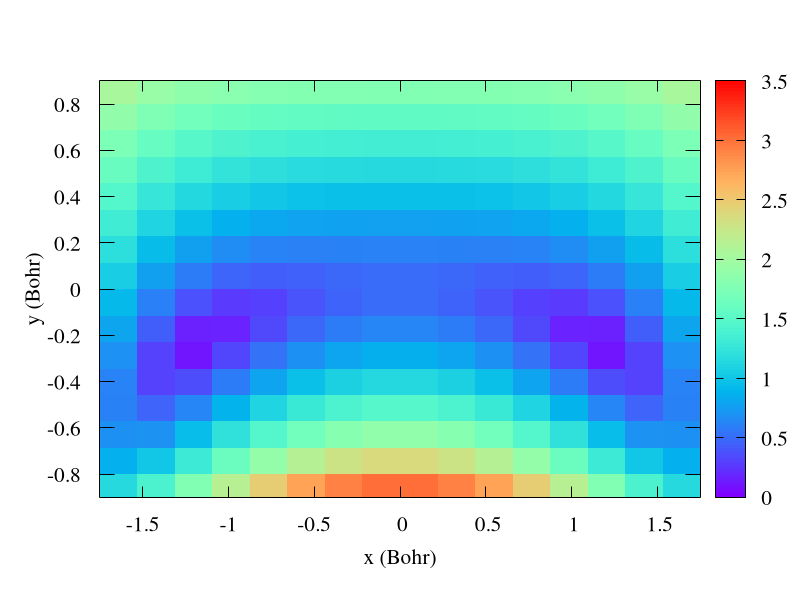}};
\node (A) at (-0.3,1.5) {\includegraphics[trim={6cm 7cm 4cm 5cm},clip,scale=0.3,angle=90]{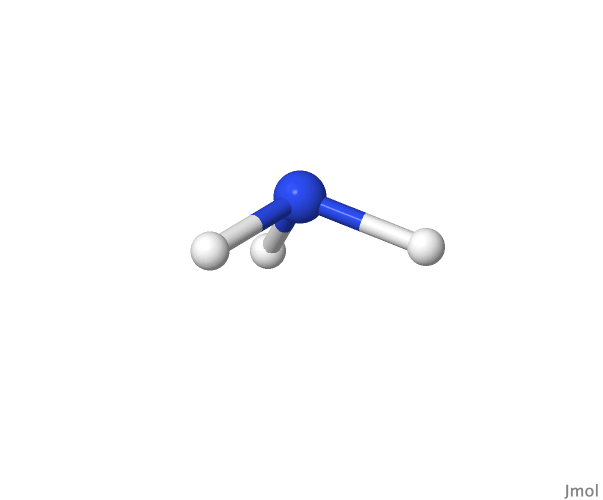}};
\node (B) at (3.8,1.4) {\includegraphics[trim={7cm 7cm 5cm 5cm},clip,scale=0.3,angle=90]{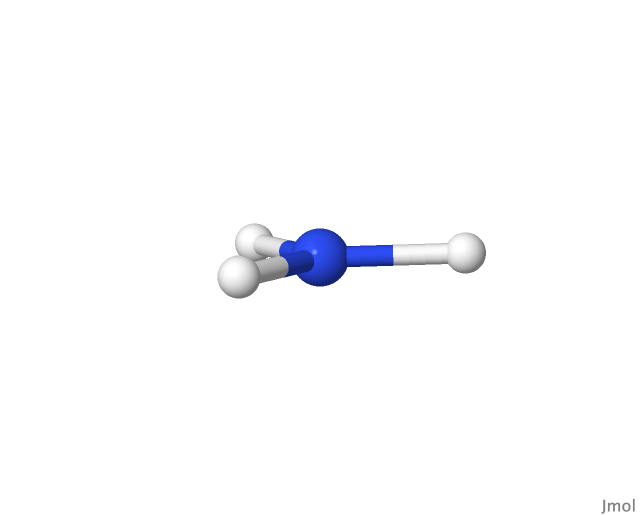}};
\node (C) at (8.5,1.5) {\reflectbox{\includegraphics[trim={6.5cm 7cm 4cm 5cm},clip,scale=0.3,angle=90]{NH3_opt.png}}};

\draw[dashed,->, ultra thick] (-0.5,0.01) -- (-0.2,-6);
\draw[dashed,->, ultra thick] (4.0,0.01) -- (4.0,-4.8);
\draw[dashed,->, ultra thick] (8.8,0.01) -- ( 8.2,-6);
\end{tikzpicture}
\caption{Energy surface of NH$_3$ with respect to the displacements along the umbrella mode (x) and symmetric stretch mode(y) on the two-dimensional Hermite grid. Note that the square root of the energy ($E(x,y)-E_{min}$) is used for the plot to enhance color variation. Therefore, the actual unit for the color map is eV$^{1/2}$.}
    \label{fig:enter-label}
\end{figure*}

NH$_3$ has a pyramidal geometry in which the nitrogen atom stays above the plane of the three H atoms. The umbrella mode of the molecule is influenced by the nitrogen atom tunneling through the plane of the hydrogen atoms to invert the pyramid\cite{dennison1932two}. The tunneling effect causes splitting of the vibrational energy levels. We utilized our coupled-anharmonic oscillator approach to simulate this effect, by allowing the umbrella and the symmetric bond stretch modes to couple. First, we found the optimal geometry of NH$_3$ when it is flat. This is known as the transition state geometry for N-tunneling reaction. Then, a harmonic vibrational analysis was performed to find normal modes and corresponding eigenvectors for the flat geometry. Hermite grid points are always symmetric around the reference geometry for a given eigenmode. By selecting the transition state geometry of NH$_3$ as the reference, we can generate Hermite grid points that allow integration of the PES along the umbrella mode, including the region containing the two minima along this mode.
 Fig.~\ref{fig:enter-label} represents the energy surface of the molecule with respect to displacements along the eigenvectors (alternatively the vibron displacement vectors) of the umbrella (x) and symmetric stretch (y) modes. This was done with a 16x16 Hermite grid. At $x=0$ (flat), the energy surface along y is a minimum at y=0, while at y=0, the energy surface along x at $x=0$ is a maximum.

\begin{figure*}[htb!]
    \centering
    \includegraphics[trim={2cm 10cm 1.8cm 2cm},clip,scale=0.8]{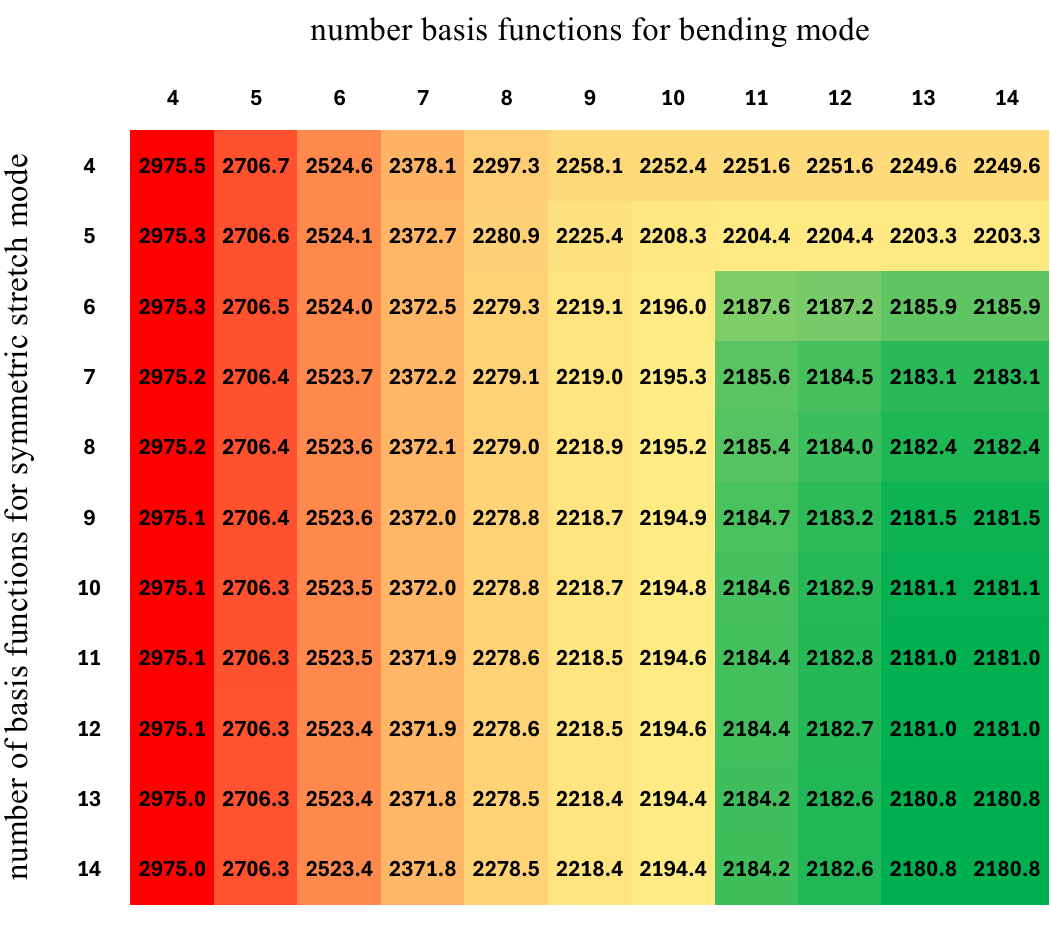}
    \caption{The ground-state vibrational energy (in cm$^{-1}$) of bending and symmetric stretch modes in NH$_3$ with different number of basis functions.}
    \label{fig:matrix_nm}
\end{figure*}

We investigated the number of basis functions required for each mode to obtain converged results. Figure \ref{fig:matrix_nm} presents the vibrational energy in the ground state when a different number of basis functions (4 to 14) is used per mode. These results confirm that the ground-state energy converges with a comparatively small number of basis functions for the symmetric stretch mode, while the bending mode requires a larger basis set to achieve convergence. 
Table \ref{tab:basis_vs_energy} shows how the ground state energy level and the splitting vary with the number of basis functions per mode. For simplicity, the same number of basis functions were used for both modes. The reported experimental splitting for the ground state energy level due to tunneling is $\sim$0.8 cm$^{-1}$,\cite{benedictvibration}. \textcolor{black}{Because of the poor convergence in the calculated splitting, we further investigated this by adding more basis functions. Hermite grids of 20x20 and 24x24 were utilized and the PES was calculated on them for the numerical integration. In both cases, basis functions up to $x^{20}e^{-A^2x^2/2}$ were added in both the umbrella and symmetric stretch modes. Then, we noticed that the calculated splitting converges to $\sim0.84$ cm$^{-1}$}.      

\begin{figure*}[htb!]
\centering
\begin{tikzpicture}
\node (D) at (4,-11) {\includegraphics[scale=0.5]{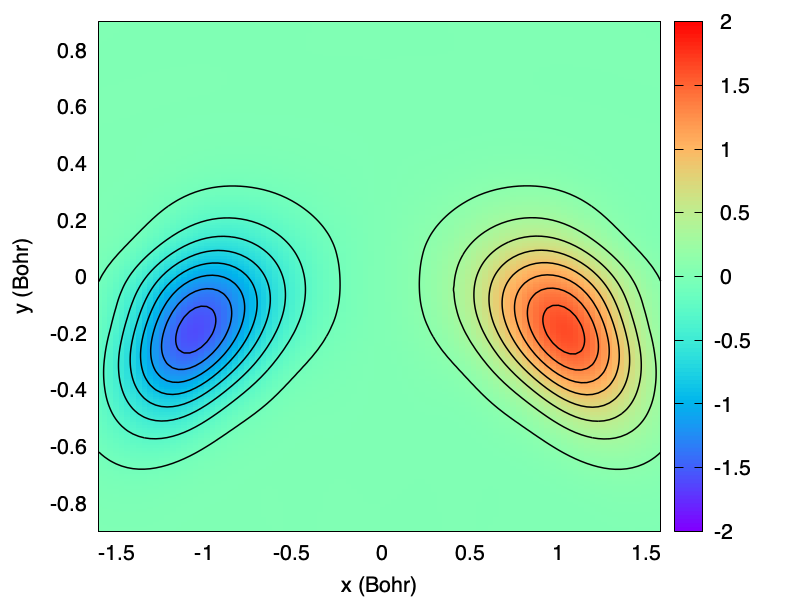}};
\node (A) at (4,0) {\includegraphics[scale=0.5]{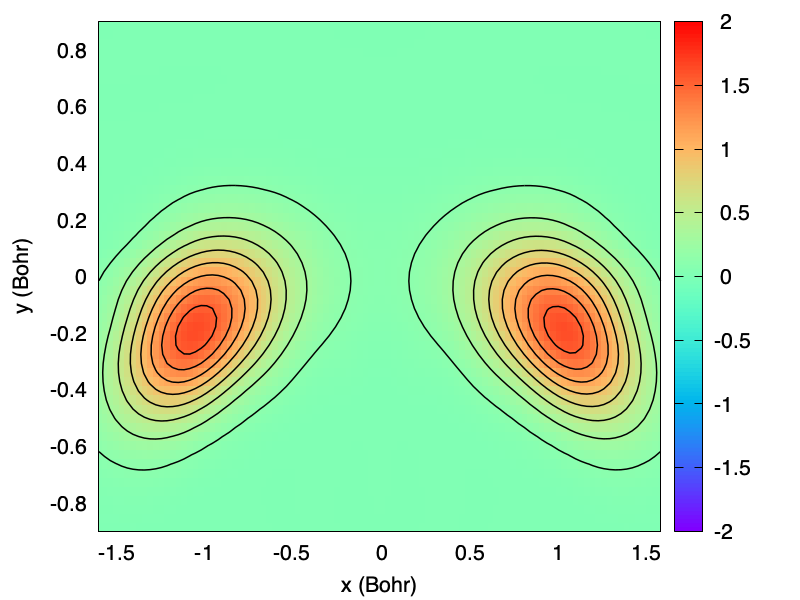}};
\end{tikzpicture}
\caption{Ground-state (top) and first excited-state (bottom) coupled-anharmonic wavefunctions for the umbrella and symmetric stretch modes were computed using 14 basis functions ($l = 0$ to $13$) per mode.
}
\label{fig:-wf}
\end{figure*}

Fig.~\ref{fig:-wf} presents the ground-state (top) and the first excited-state (bottom) wavefunctions corresponding to the coupled anharmonic oscillators for the umbrella and symmetric stretch vibrational modes of NH$_3$. From Fig.~\ref{fig:-wf}, it is clear that the ground-state wavefunction is symmetric around x=0 along the x-axis, while the first excited state exhibits anti-symmetric behavior. 

\begin{table*}[htb!]
    \begin{tabular}{crrr}
    \toprule
\begin{tabular}[c]{@{}c@{}}Number of basis \\ functions per mode\end{tabular} 
& E$_0$ (cm$^{-1}$) 
& E$_1$--E$_0$ (cm$^{-1}$)& E$_2$--E$_0$ (cm$^{-1}$) \\
\midrule
2  & 3601.69 & 78.75  & 3533.31  \\
3  & 3171.43 & 508.05 & 1078.75  \\
4  & 2975.54 & 194.11 & 1272.53  \\
5  & 2706.63 & 268.68 &  991.93  \\
6  & 2523.98 & 182.56 & 1174.57  \\
7  & 2372.19 & 151.55 & 1041.60  \\
8  & 2278.96 &  93.13 & 1134.75 \\
9  & 2218.72 & 60.08  &  977.92 \\
10 & 2194.78 & 23.89  &  990.52 \\
11 & 2184.42 & 10.19  &  905.38 \\
12 & 2182.72 & 1.62   &  906.95 \\
13 & 2180.84 & 1.76   &  890.16\\
14 & 2180.80 & 0.35   &  890.10\\
\bottomrule
    \end{tabular}
    \caption{Dependence of ground-state, first two excited-state energies relative to ground state on the number of basis functions per mode for NH$_3$}
    \label{tab:basis_vs_energy}
\end{table*}

It is worth discussing how the exponential factor for the Gaussian basis functions ($A^2$) was chosen for the inversion mode.  This mode features a local potential energy maximum at x = 0, leading to an imaginary frequency in the harmonic approximation. However, when coupling this mode with the symmetric stretch mode and accounting for anharmonic effects, the frequency becomes real.  As a pragmatic choice, we used the magnitude of the imaginary frequency to define the exponential factor as $A^2=|\omega| M_A$, where $M_A$ is the effective mass. To assess the sensitivity of our results to this choice, we varied  $A^2$ across a range extending from 50\% below to 50\% above the value based on the imaginary frequency.  
Each variation requires redefining the Hermite-grid and recomputing the potential energy surface. The corresponding $A^2$ values and resulting ground-state energies are summarized in Table~\ref{expne}. Our findings demonstrate that the ground-state energy exhibits only a weak dependence on the specific value of $A^2$ used.  
\begin{table*}[htb!]
    \begin{tabular}{c|c}
    \toprule
    Exponential factor (Bohr$^{-2}$)& Ground-state energy (cm$^{-1}$)\\
    \midrule
2.042(-50.0\%) &	2207.13(1.21\%) \\
3.268(-20.0\%) &	2186.20(0.25\%) \\
3.676(-10.0\%) &	2182.46(0.08\%)  \\
4.085(0.0\%)   &	2180.80(0.00\%)  \\
4.493(10.0\%)	 & 2178.16(-0.12\%)  \\
\textbf{5.130(25.6\%))}	 & \textbf{2175.95(-0.22\%)}  \\
5.310(30.0\%)	 & 2174.96(-0.27\%)  \\
5.718(40.0\%)	 & 2182.32(0.07\%)   \\
6.127(50.0\%)	 & 2197.40(0.76\%) \\
\bottomrule
    \end{tabular}
    \caption{Dependence of the ground-state energy ($E_0$)of the coupled-anharmonic oscillator on the exponential ($A^2$) factor used in the Gaussian basis functions for the inversion mode of NH$_3$. The numbers in the brackets represent the percentage deviation 
    relative to the ones obtained 
    using the value of $A^2$ based on the imaginary frequency of the harmonic calculation for the flat NH$_3$ structure.  The $A^2$ value in bold corresponds to that derived from the harmonic approximation of the optimal bent NH$_3$ structure.} 
    \label{expne}
\end{table*}

\subsubsection{CO$_2$: Fermi-resonance}
Vibration of the CO$_2$ molecule is a classic example where anharmonicity leads to an interesting phenomenon referred to as the Fermi-resonance\cite{fermi1931ramaneffekt}. CO$_2$ is a linear molecule with 4 vibrational modes. The first overtone frequency of the bending mode (which is doubly-degenerate) happens to be approximately equal to the frequency of the symmetric stretch mode. Both have the same symmetry. Since the actual vibrational potential energy has anharmonic contributions,  they interact with each other and produce mixed and split vibrational states. As a result, the vibrational spectrum of CO$_2$ shows additional Raman active modes. This is known as the  Fermi resonance in CO$_2$.   

\begin{figure*}[htb!]
\begin{tikzpicture}[xshift=2cm]
\node (A) at (0,0) {\includegraphics[scale=0.35,angle=0]{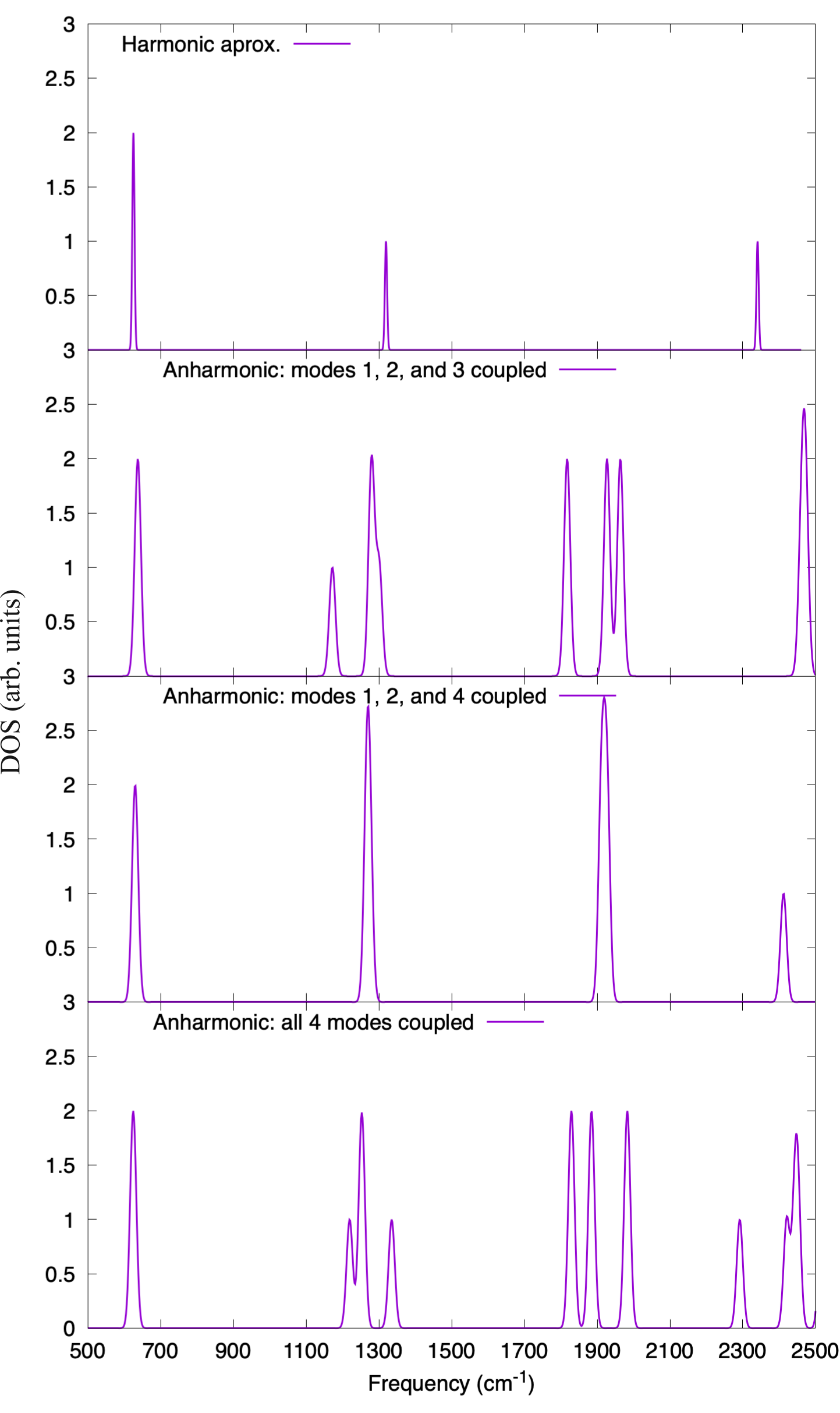}};
\node (a) [text width=10cm] at (0.1,9.5) {625};
\node[text width=10cm] at (4.5,7.5) {1319 };
\node[text width=10cm] at (9.9,7.8) {2342};
\node[text width=10cm] at (0.1,4.3) {637};
\node[text width=10cm] at (2.9,2.6) {1172};
\node[text width=10cm] at (4,4.4) {1280};
\node[text width=10cm] at (4.4,2.5) {1300};
\node[text width=10cm] at (6.3,4) {1817};
\node[text width=10cm] at (7.4,4.5) {1927};
\node[text width=10cm] at (8.3,4) {1964};
\node[text width=10cm] at (10.3,5.2) {2468};
\node[text width=10cm] at (0.1,-1) {630};
\node[text width=10cm] at (3.2,-0.5) {1270};
\node[text width=10cm] at (6.9,-0.5) {1919};
\node[text width=10cm] at (10.2,-2.6) {2412};
\node[text width=10cm] at (0.1,-6.2) {625};
\node[text width=10cm] at (2.9,-8) {1219};
\node[text width=10cm] at (3.5,-6.2) {1253};
\node[text width=10cm] at (4.4,-7.9) {1334};
\node[text width=10cm] at (6.5,-6.9) {1830};
\node[text width=10cm] at (7.3,-6.1) {1885};
\node[text width=10cm] at (8.4,-6.8) {1982};
\node[text width=10cm] at (9.1,-8.1) {2292};
\node[text width=10cm] at (10.1,-7.8) {2421};
\node[text width=10cm] at (10.4,-6.5) {2447};
\draw[dashed,line width=2pt] (-0.87,5.7) -- (-0.87,7.4) ;
\node (b) [text width=10cm] at (2.3,7.5) {1250(x3)};

\draw[dashed,line width=2pt] (2.7,5.7) -- (2.7,7.4) ;
\node (c) [text width=10cm] at (5.9,7.2) {1875(x4)};
\draw[dashed,line width=2pt] (3.1,5.7) -- (3.1,7.4) ;
\node (d) [text width=10cm] at (8.15,7.2) {1944(x2)};

\node (e) [text width=10cm] at (3,9) {+625};
\node (f) [text width=10cm] at (4.2,9.8) {+1250};
\node (g) [text width=10cm] at (5.3,8.3) {+625};

  \node (A) at (-4.3,9.5) {};
  \node (B) at (-0.9,7.4) {};
  \node (C) at (2.7,7.4) {};
  \node (D) at (3.1, 7.4) {};
  \node (E) at (-0.6, 7.6) {};
\draw[->, bend left, very thick] (A) to (B);
\draw[->, bend left, very thick] (A) to (C);
\draw[->, bend left, very thick] (E) to (D);

\node[text width=10cm] at (9,10) {(a)};
\node[text width=10cm] at (9,4.7) {(b)};
\node[text width=10cm] at (9,-0.5) {(c)};
\node[text width=10cm] at (9,-5.7) {(d)};

\end{tikzpicture}

\caption{Calculated vibrational density of states (VDOS) of CO$_2$ molecule. (a) VDOS from the Harmonic approximation. Dashed lines represent the next excited states. (b) VDOS from an anharmonic calculation when the first 3 modes are allowed to couple. (c) VDOS from an anharmonic calculation when the first 2 and the last mode are allowed to couple. (d) VDOS from an anharmonic calculation when all 4 modes are allowed to couple. }
    \label{CO2_DOS_4}
\end{figure*}

We modeled this phenomenon using our coupled-anharmonic oscillator approach. First, the geometry of CO$_2$ molecule was optimized. Then a harmonic analysis was performed. From the harmonic approximation, the first two vibrational modes, at 625 cm$^{-1}$, correspond to the bending modes. The third mode, at 1319 cm$^{-1}$, corresponds to the symmetric stretch mode, while the last mode, at 2342 cm$^{-1}$, corresponds to the asymmetric stretch mode.  As described earlier,  the harmonic frequencies and the effective masses of four modes from the harmonic analysis were used to construct the basis functions for the coupled-anharmonic Hamiltonian. We studied the effect of anharmonicity and coupling between modes in CO$_2$ in three different ways. Firstly, only the first three modes were coupled. Secondly, the first two and the fourth modes were coupled. Finally, all four modes were allowed to couple. Three different multi-dimensional Hermite-quadrature grids were defined based on the three cases mentioned above. Convergence tests were performed that involved 6 and 8 points per mode. The number of potential energy evaluations required is $8^m$ where $m$ is the number of modes that are allowed to couple. In Fig.~\ref{CO2_DOS_4}, we present the calculated vibrational density of states (DOS) from the harmonic approximation and the anharmonic calculations when different modes are allowed to couple. 

In Fig.~\ref{CO2_DOS_4}(a), the vibrational density of the states from the harmonic approximation (purple) is presented. The next excited states of the bending modes are indicated using black dashed lines. Because there is no mixing between modes within the harmonic approximation, no additional peaks appear. However, when the bending modes and the symmetric stretch mode were treated anharmonically, the modes are coupled and extra peaks start to appear (Fig.~\ref{CO2_DOS_4}(b)). Here, we focus on the region of the symmetric stretch mode. For this case, three peaks (instead of two) are visible in the region corresponding to the symmetric stretch. This is because of the splitting caused by mixing the symmetric stretch mode and the excited state of the bending modes. This behavior is the Fermi-resonance. Fig.~\ref{CO2_DOS_4} (c) shows only sharp peaks that correspond to the first modes and there is no evidence of coupling between the bending and asymmetric stretch modes (Fig.~\ref{CO2_DOS_4} (c)). Fig.~\ref{CO2_DOS_4} (d) presents the vibrational DOS when all four modes were treated anharmonically and allowed to couple.

\begin{figure*}[!htb]
\begin{tikzpicture}[xshift=2cm]
  \node (A) at (0,0) {
    \includegraphics[scale=0.4]{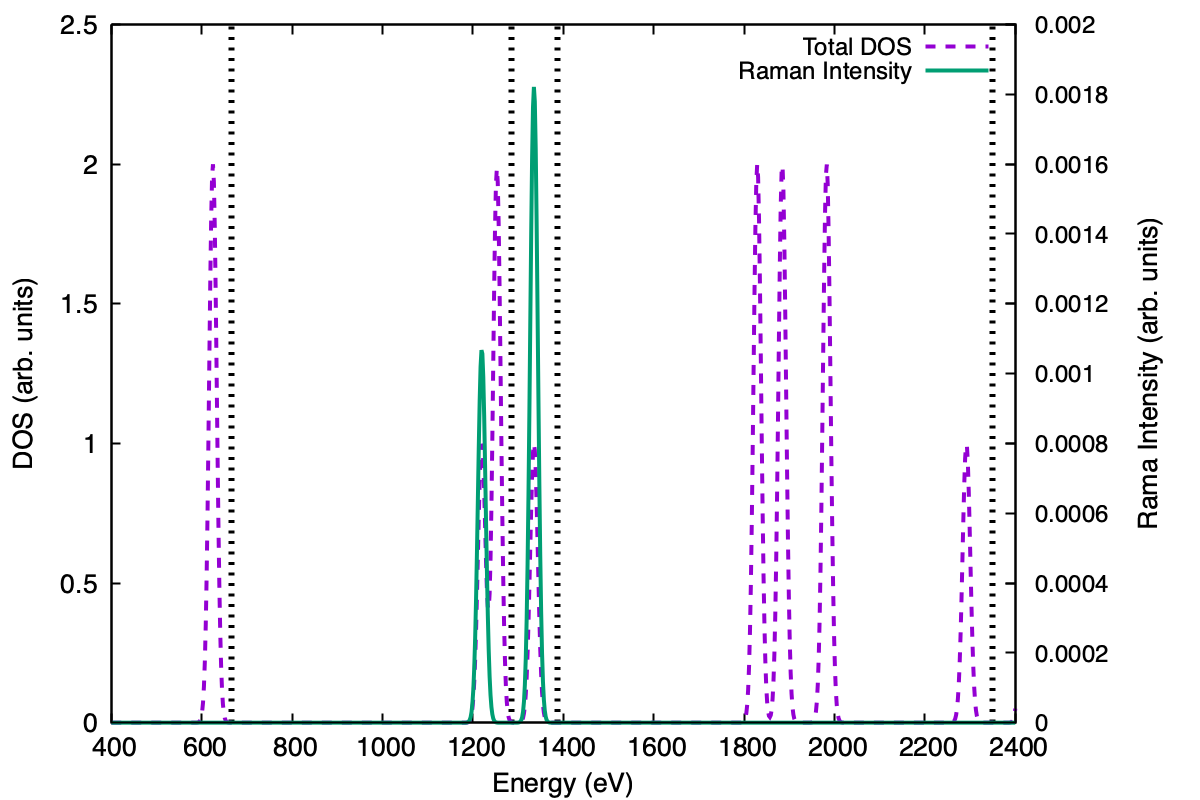}};
\node[text width=10cm,rotate=90] at (-5,7) {667};
\node[text width=10cm,rotate=90] at (-1.5,9) {1285};
\node[text width=10cm,rotate=90] at (-0.4,9) {1388};
\node[text width=10cm,rotate=90] at (5.3,7) {2349};

    \end{tikzpicture}
       \caption{Calculated vibrational density of states and approximate Raman  activity of the CO$_2$ molecule when all four modes are allowed to couple. \textcolor{black}{The experimental peaks are indicated using vertical dashed lines in black.}} 
         \label{CO2_DOS}
\end{figure*}

 Fig.~\ref{CO2_DOS} shows the vibrational DOS and the Raman active modes. As expected from the Fermi resonance, the two modes at 1219 cm$^{-1}$ and 1334 cm$^{-1}$ are Raman active. Although these values do not exactly match the experimental observations\cite{rasetti1929raman,Dickinson1929,herzberg2013molecular} (1285 cm$^{-1}$ and 1388 cm$^{-1}$), the splitting between the two modes obtained from our calculation (115 cm$^{-1}$) is in close agreement with the experimental value (103 cm$^{-1}$), indicating that our model captures the essence of the anharmonic coupling responsible for the Fermi-resonance in CO$_2$.

\section{Summary}
In this work, we presented a framework for implementing the VCI method based on a Gaussian function basis set.  Owing to this choice of basis set, overlap matrix elements, kinetic energy terms in the Hamiltonian, and properties such as moments can be evaluated analytically. Only for the PES integration do we employ an efficient numerical grid derived from the Gauss–Hermite quadrature method. The PES were obtained from DFT total energy calculations using the GGA-PBE xc functional. As an initial application, we tested our approach on one-dimensional cases (diatomic molecules) and found that converged vibrational spectra can be obtained with as few as six basis functions. We also found that the corrections to bond lengths due to anharmonicity in these diatomic molecules were important when their constituents are H atom and alkali or alkaline earth metal atoms. We further demonstrated that the method can be generalized to model coupled anharmonic oscillations in molecules containing more than 2 atoms. We allowed the symmetric stretch and umbrella modes to couple in NH$_3$ molecule. By doing that, we were able to demonstrate energy level splitting due to N-tunneling through the plane of the hydrogen atoms. \textcolor{black}{However due to the complexity of the PES along the umbrella mode, higher order basis functions are required to achieve  convergence in both the ground-state energy and and splitting. This requires a denser Hermite grid for accurate PES integration. We view this behavior as a limitation of the method}.  In the CO$_2$ molecule, we showed that additional modes (splitting) occur only when the two bending modes and the symmetric stretch modes were allowed to couple. In the case where we allowed all four modes to couple, we found that there were two Raman active modes at 1219 cm$^{-1}$ and 1334 cm$^{-1}$. This behavior is known as Fermi-resonance. These examples demonstrate that our approach captures important physical phenomena that modifies the vibrational spectrum of molecules due to anharmonicity in the PES.

\textcolor{black}{The cost of a full harmonic vibrational calculation when performed thru a series of SCF calculations is proportional to $T_{harm}=(6N)\times I_e \times N^3 \times B^3$+$(3N)^3$.  
   Here N is the number of atoms and B represents the number of basis functions per atom. The number of iterations required per displacement ($I_e$) is approximate 6 since wavefunctions from the previous displacement are an excellent starting point for the next displacement.
   The cost of a full harmonic vibrational calculation followed by a {\bf complete} anharmonic analysis, on a limited number ($N_A\le3N$) of the harmonic modes is given by: $T_{anh}^{full}=T_{harm}+(2\sum_{i=1}^{N_A}b^i_A)\times I_e \times N^3 \times B^3+[\Pi_{i=1}^{N_A}b^i_A]^3$. Here, $b_i$ is at least 2 and represents the mesh size needed for the gaussian Hermite grid (which for the case of NH$_3$ turned out to be 24. For this case, it is not the extra number of SCF calculations that is the rate determining step. The rate determining step becomes the last term $[\Pi_{i=1}^{N_A}b^i_A]^3$ as it is significantly greater than $2^{3N_A}$. So, for example if one chose the seemingly small number of modes ($N_A=10$), the cost of the vibrational diagonalization using our approach would be $2^{30}$ which is obviously awful scaling (AS). What our approach can allow us to eventually do is find a smaller number of partner modes, by identifying nearly degenerate pairs of higher harmonics, for each of the $N_A$ interesting degrees of vibrational freedom (maybe 3-4) and then perform a series of {\bf incomplete} low-scaling anharmonic diagonalizaton on each of the $N_A$ modes and then use a small number of states from each of these calculations to perform a final full-scale anharmonic calculation at a diagonalization cost that scales as $N^3$. Many details needed to be attended to to make this happen. However, we note above identification of the limited number ($N_A$) of vibrational modes requires excellent chemical intuition or the initial harmonic analysis but the resulting complexity of a NEO-like analysis of a multi-dimensional vibrational space would have the same AS as our complete method albeit with an amplifying prefactor of $I_e$. Perhaps there is a similar incomplete method for coupled methods but these details also need to be worked out. Multiple methods are needed to (1) reduce $b_i$, for which there are prescriptions, and to automatically and systematically identify the interesting vibrational degrees of freedom.}

  \textcolor{black}{Anharmonic vibrational effects can be treated at either the Born-Oppenheimer approximation, at the level of configuration interaction, or in more complicated cases the well established nuclear-electronic orbital method in which case both electrons and nuclei are treated quantum mechanically. Dating back to early work by theoretical work by Goddard~\cite{SURRATT197739} (including references therein) and experimental work by Herzberg
    \cite{Herzberg,Herzber46}, 
    there is a history of work on the methyl (CH$_3$) radical. From computational and theoretical perspectives Born-Oppenheimer effects and multi-configurational effects have been examined in attempting to understand the double- versus single- well approximations in this case. An analysis of that series of papers suggests that one can not necessarily criticize a lower-level approach which yields a Born-Oppenheimer double well just because a higher level predicts a single well and instead one must look at the reliability of the predicted vibrational spectra. The reason for this is that total energy, which includes purely electronic and anharmonic vibrational terms in the former case compared to a single total energy in the second case leads to the fact that the resulting energy vs displacement curves do not distinguish between standard zero-point contributions and electronic contributions so the appearance of single-well nature in the total energy does not contradict BO results. This fact has been overlooked in more recent cases that argue for spurious symmetry breaking for a new density functional method for the case of the methyl radical~\cite{Hahn}. Perhaps  because of such issues, Hammes$-$Schiffer and others have argued for explicit product functions that include the nuclear and electronic wavefunction~\cite{NEOSHS,SHSXLi} and there are of course countless other examples where gaussians have been used for multi-center basis functions.  At one level one could argue that an explicitly self-consistent product representation is better than Born-Oppenheimer approaches as it adds nuclear quantum effects in a manner that does not significantly change scaling vs system size and only adds the appearance of a diagonalization of the standard electronic Hamiltonians, for both spins, that is performed at the same time as a much smaller solution to the vibrational wavefunctions. However effectively avoiding non-systematic errors due to the electron-nuclear cusp condition~\cite{CuspRJNeeds} is guaranteed in a Born-Oppenheimer based approach which renders a plug-and-play software for to all electronic-structure methods - particularily all-electron software.  However, in regard to any method describing vibrational effects of nuclei, it is the case that when one is interested in accounting for full anharmonicity involving many nuclear degress of freedom, the time required for simultaneous diagonalization of the vibrational quickly accelerates O($M^3$)O($N^3$), with M and N the number of basis functions per mode and atoms respectively. For the general case there are not {\em a priori} approaches for conjuring up which modes, to select from the 3N-possible phonon eigenstates,  to address more perfectly. The method proposed determines the interesting anharmonic/vibrational space from the initial vibrational calculation~\cite{porezag}. Here we argue that a post-SCF anharmonic calculation which allows for both identification and treatment of chemically interesting anharmonic couplings between nearly degenerate albeit distinct vibronic states could be coupled to many approaches include NEO. We hope that future studies will use the approach described here to stray further away from the use of standard Hermite polynomials and the resulting Gauss-Hermite-Quadrature approaches to a prescription that instead uses the the resulting space of vibrational eigenstates determined for the one-dimensional space to generate reduced grids and reduced basis functions. }
\section*{Supplementary Material}
See the Supplementary Material for the reference geometries, normal-mode eigenvectors, Gauss–Hermite grid points, and corresponding weights for NH$_3$ and CO$_2$ used in the anharmonic analyses.

\begin{acknowledgments}
The calculations were performed on the Jakar high-performance computing system provided by UTEP. J.P., M.I., and M.R.P. were supported by the Tec$^4$ project, which is funded by the U.S. Department of Energy, Office of Science, Office of Basic Energy Sciences, the Division of Chemical Sciences, Geosciences, and Biosciences (under Grant No. FWP 82037). K.P.K.W. and K.A.J. were supported by the FLOSIC project which is funded by the U.S. Department of Energy, Office of Science,
Office of Basic Energy Sciences, as part of the Computational Chemical Sciences Program (under Award No. DE-SC0018331).
\end{acknowledgments}

\section*{Data Availability Statement}

The data that support the findings of
this study are available within the
article and its supplementary material.

\bibliography{aipsamp}

\end{document}